\documentclass[letterpaper]{article} 
\usepackage{sty/aaai25}  
\usepackage{times}  
\usepackage{helvet}  
\usepackage{courier}  
\usepackage[hyphens]{url}  
\usepackage{graphicx} 
\usepackage{natbib}  
\usepackage{caption} 
\usepackage{graphicx}
\usepackage{url}
\usepackage{booktabs}
\usepackage{multirow}
\usepackage{amsmath,amssymb}
\usepackage{algorithm}
\usepackage[noEnd=true,indLines=true,commentColor=black]{sty/algpseudocodex}
\usepackage{siunitx}
\usepackage{cleveref}
\usepackage{sty/custom}

\usepackage[subrefformat=parens]{subcaption}
\usepackage{newfloat}
\usepackage{listings}
\DeclareCaptionStyle{ruled}{labelfont=normalfont,labelsep=colon,strut=off} 
\floatstyle{ruled}
\newfloat{listing}{tb}{lst}{}
\floatname{listing}{Listing}
\makeatletter
\gdef\copyright@on{}
\makeatother

\title{Lightweight and Effective Preference Construction in PIBT\\for Large-Scale Multi-Agent Pathfinding}
\author{Keisuke Okumura\textsuperscript{\rm 1,\rm 2}, Hiroki Nagai\textsuperscript{\rm 1,\rm 3}}
\affiliations{
\textsuperscript{\rm 1}National Institute of Advanced Industrial Science and Technology (AIST), Japan\\
\textsuperscript{\rm 2}University of Cambridge, UK\\
\textsuperscript{\rm 3}Keio University, Japan\\
\{okumura.k, nagai.hiroki39 \}@aist.go.jp
}

\begin{document}

\maketitle
\begin{abstract}
PIBT is a computationally lightweight algorithm that can be applied to a variety of multi-agent pathfinding (MAPF) problems, generating the next collision-free locations of agents given another.
Because of its simplicity and scalability, it is becoming a popular underlying scheme for recent large-scale MAPF methods involving several hundreds or thousands of agents.
Vanilla PIBT makes agents behave greedily towards their assigned goals, while agents typically have multiple best actions, since the graph shortest path is not always unique.
Consequently, tiebreaking about how to choose between these actions significantly affects resulting solutions.
This paper studies two simple yet effective techniques for tiebreaking in PIBT, without compromising its computational advantage.
The first technique allows an agent to intelligently dodge another, taking into account whether each action will hinder the progress of the next timestep.
The second technique is to learn, through multiple PIBT runs, how an action causes regret in others and to use this information to minimise regret collectively.
Our empirical results demonstrate that these techniques can reduce the solution cost of one-shot MAPF and improve the throughput of lifelong MAPF.
For instance, in densely populated one-shot cases, the combined use of these tiebreaks achieves improvements of around 10-20\% in sum-of-costs, without significantly compromising the speed of a PIBT-based planner.
\end{abstract}

\section{Introduction}
Multi-agent pathfinding (MAPF)~\cite{stern2019def} is a planning problem that assigns collision-free paths to each agent on a discrete workspace, and is seen as a future driving force of efficient logistics automation~\cite{ma2017lifelong,li2021lifelong}.
Although real-world applications often require planning algorithms to cope with swarms of hundreds or thousands of robots~\cite{amazon2022robots}, synthesising optimal coordination is computationally challenging~\cite{yu2013structure,banfi2017intractability}.
It is therefore important to develop suboptimal methods that are scalable, but can produce sufficiently plausible solutions with fewer redundant agent movements within tight time constraints.
Indeed, developments in suboptimal MAPF methods over the last decade have been remarkable, allowing us to obtain collision-free paths for hundreds, sometimes thousands, of agents in sub-seconds~\cite{li2022mapf,okumura2023lacam}.

Among advanced suboptimal methods, PIBT (Priority Inheritance with Backtracking)~\cite{okumura2022priority} has gained notable popularity in recent years.
It is a computationally lightweight function that generates a collision-free, transitionable \emph{configuration}, i.e., locations for all agents, given another.
Due to its simplicity and scalability, PIBT has been serving as a core component in state-of-the-art MAPF studies, such as massively scalable search-based algorithms with theoretical guarantees~\cite{okumura2023lacam,okumura2023lacam2}, a winning strategy~\cite{jiang2024scaling} of the first MAPF competition held in 2023~\cite{chan2024league}, a lifelong MAPF approach with hundreds of times faster runtime than a search-based method with comparable solution quality~\cite{zhang2024guidance}, and machine learning approaches that can handle densely populated instances, which are traditionally difficult~\cite{veerapaneni2024improving,jiang2024deploying}.
In this sense, the improvement of the PIBT scheme has a significant impact on practical solutions for large-scale MAPF systems.

Implementation-wise, for each configuration generation PIBT uses a sorted list of candidate actions for each agent, herein termed \emph{preference}.
For an agent $i$ within a graph $G=(V, E)$ at a vertex $u \in V$, an action corresponds to a neighbouring vertex $v$, including $u$ itself, which represents `staying' motion.
These candidates are then typically sorted by the shortest path distance on $G$, denoted as \dist, to a target vertex $g_i \in V$.
Formally, they are sorted in a lexicographic and ascending order with
\begin{equation}
  \left\langle \dist(v, g_i), \epsilon \right\rangle
  \label{eq:preference}
\end{equation}
where $\epsilon$ is a random number to break ties.

This preference construction ensures that agents move towards their respective destinations, but it remains a bare minimum implementation.
In fact, it is well known in the research community that \cref{eq:preference} leads to greedy and shortsighted suboptimal behaviour, as evidenced by recent studies~\cite{okumura2024lacam3,chen2024traffic,veerapaneni2024improving,zhang2024guidance,zang2025online} that improve PIBT preference in terms of solution quality.
These studies can improve on the original PIBT, but at a significant computational cost from the original, which impairs the advantage of being able to instantly generate motions for thousands of agents or more.

Instead, we are interested in computationally lightweight yet effective preference construction in PIBT.
In particular, this paper proposes to add two terms before the random tiebreaker, as:
\begin{equation}
\left\langle \dist(v, g_i), \hindrance, \regret, \epsilon \right\rangle
\label{eq:hindrance-regret}
\end{equation}
These interim terms are efficiently computed with an overhead no greater than the linear time complexity of PIBT itself, and therefore do not compromise the algorithm scalability.
Nevertheless, this enhanced tiebreaking significantly improves the solution quality in MAPF problems.
For example, in densely populated one-shot MAPF scenarios, we observe that a PIBT-based planner (i.e., LaCAM) with \cref{eq:hindrance-regret} achieves solution cost reductions of around 10-20\% over \cref{eq:preference}.
Furthermore, in a specific lifelong case with a map size of $32\times 32$, including 10\% obstacles and 400 agents, \cref{eq:hindrance-regret} contributes to $\geq 40\%$ throughput improvement over vanilla PIBT.
This empirical evidence suggests that the enhanced tiebreaking may improve many state-of-the-art MAPF implementations based on PIBT with little engineering effort.

Intuitively, the first term, \hindrance, is a one-step later estimate of whether an action hinders the progress of a neighbouring agent towards its goal, computed by $O(\Delta(G))$, where $\Delta(G)$ is the maximum degree of a graph $G$.
The second term, \regret, represents how an action affects the subsequent action choices of nearby agents within that configuration generation, which is learned by running PIBT a few times.
The overhead is thus the time complexity of PIBT itself, which is practically sub-millisecond for each configuration generation, even with thousands of agents.
With this combination, we make a minimal but effective effort to improve PIBT.
In addition, the completeness guarantees in the original PIBT and LaCAM are maintained, as the proposed adjustments are just tiebreaking.

In what follows, the paper continues with the preliminaries, related work, technical details, empirical results and discussions in order.

\section{Preliminaries}

\subsection{Problem Definition}
This paper addresses both one-shot and lifelong versions with a classical MAPF fashion.
The system consists of a set of agents $A = \{1, 2, \ldots, n\}$ and a graph $G=(V, E)$.
All agents act synchronously according to a discrete wallclock time.
At each timestep, each agent can stay at its current vertex or move to a neighbouring vertex.
Vertex and edge conflicts are considered, i.e. two agents cannot occupy the same vertex simultaneously, and two agents cannot swap their occupied vertices within one timestep move.
Then, a \emph{one-shot MAPF problem} aims to find a finite action sequence for each agent $i \in A$ to its assigned goal $g_i \in V$ from a given start $s_i \in V$.
The solution quality is assessed by \emph{Sum-of-Costs (SoC)} (aka. flowtime), which sums the travel time of each agent until it stops at the target location and no longer moves.
In a \emph{lifelong MAPF problem}, once an agent has reached its goal, a new goal is immediately assigned by an external, black-box task assigner, so planners need to operate the agents continuously.
The solution quality is assessed by \emph{throughput}, the number of task completions normalised by the operation period.

\subsection{PIBT}
PIBT is a function that maps a collision-free configuration $\Q\from \in V^{|A|}$ to another $\Q\to$.
These configurations are ensured to be transitionable in the sense that each agent $i$ can move from the current location $\Q\from[i]$ to $\Q\to[i]$ within one timestep move, without colliding.

\Cref{algo:pibt} provides the pseudocode.
The assignment process is performed by calling a submodule (Lines~\ref{algo:pibt:recursive:start}--\ref{algo:pibt:recursive:end}) sequentially following a prefixed order of agents $A$ (\cref{algo:pibt:call-func}).
PIBT then tries to assign each agent a location in order of its preference $\P$, a sorted list of available actions (\cref{algo:pibt:sort}) consisting of the currently occupying vertex and its neighbouring vertices, denoted as \neigh.
Each assignment is accompanied by a collision check (\cref{algo:pibt:collision-check}), ensuring a collision-free outcome.
The agent order is dynamically adjusted when one agent $j$ blocks the desired location of another agent $i$, which is implemented by a recursive call of the submodule (\cref{algo:pibt:check-k}).
This scheme is called \emph{priority inheritance} as assignment priority is inherited from $i$ to $j$.
The priority inheritance call is followed by \emph{backtracking} (\cref{algo:pibt:call-pi}), which notifies $i$ whether $j$ secures a feasible action;
otherwise, $i$ needs the reassignment and thus continues the for-loop operation.
With this simple procedure, PIBT serves as an instant configuration generator.

Applying PIBT to either one-shot or lifelong MAPF is straightforward;
repeat one-timestep planning with a preference creation using \cref{eq:preference} with the current goal assignment.
In addition, there is a search-based method called LaCAM~\cite{okumura2023lacam} for one-shot MAPF, which uses PIBT as a successor generation.
The experiments use LaCAM for one-shot MAPF, and bare PIBT for lifelong scenarios.

{
  \begin{algorithm}[t!]
    \caption{Original PIBT}
    \label{algo:pibt}
    \begin{algorithmic}[1]
      \small
      \Input{configuration $\Q\from$, agents $A$}
      \Output{configuration $\Q\to$ (each initialized with $\bot$)}
      \State \textbf{for}~$i \in A$~\textbf{do};~
             \textbf{if}~{$\Q\to[i] = \bot$}~\textbf{then}~{$\PIBT(i)$}
      \label{algo:pibt:call-func}
      \State \Return $\Q\to$
      \label{algo:pibt:top:end}
      \item[]
      \Procedure{\PIBT}{$i$}
      \label{algo:pibt:recursive:start}
      \State $\P \leftarrow \neigh\left(\Q\from[i]\right) \cup \left\{\Q\from[i]\right\}$
      \label{algo:pibt:cand}
      \State sort $\P$ by some means
      \Comment{preference construction}
      \label{algo:pibt:sort}
      \For{$v \in \P$}
      \label{algo:pibt:loop-start}
      \IfSingle{collisions in $\Q\to$ supposing $\Q\to[i]=v$}{\Continue}
      \label{algo:pibt:collision-check}
      \State $\Q\to[i] \leftarrow v$
      \label{algo:pibt:reserve}
      \If{$\exists j \in A~\text{s.t.}~j \neq i \land \Q\from[j]=v \land \Q\to[j]=\bot$}
      \label{algo:pibt:check-k}
      \IfSingle{$\PIBT(j)= \invalid$}{\Continue}
      \label{algo:pibt:call-pi}
      \EndIf
      \State \Return~\valid
      \label{algo:pibt:valid}
      \EndFor
      \label{algo:pibt:loop-end}
      \State $\Q\to[i] \leftarrow \Q\from[i]$;~\Return~\invalid
      \label{algo:pibt:stay}
      \EndProcedure
      \label{algo:pibt:recursive:end}
    \end{algorithmic}
  \end{algorithm}
}

\subsection{Related Work}
The importance of tiebreaking in preference construction was briefly mentioned in the original PIBT paper~\cite{okumura2022priority}, but not investigated in depth.
As PIBT has grown in popularity, researchers have investigated how to optimise the preference.

An effective scheme to improve solution quality takes the form of preparing a \emph{guidance} graph~\cite{zhang2024guidance}, which is typically represented by a weighted directed graph $G_w$ over the original graph $G$.
Instead of the naive preference on $G$, such a study makes PIBT follow a preference based on $G_w$.
With carefully designed weights, either by handcrafted~\cite{cohen2015feasibility,li2023study}, computationally-heavy heuristic search~\cite{okumura2024lacam3,chen2024traffic}, or machine learning approaches~\cite{yu2023congestion,zhang2024guidance,zang2025online}, the guidance is able to create \emph{global} traffic flow and mitigate congestion.
The techniques presented in this paper are orthogonal to this notion, which aims to minimise ineffective \emph{local} collective behaviour, and thus can co-enhance PIBT with the global guiadance.

Another direction of preference optimisation focuses on overcoming narrow corridor scenarios where two agents cannot exchange their positions.
Such situations require long-horizon coordinated behaviour between paired agents, which cannot be synthesised by greedy and shortsighted PIBT.
Several studies have successfully migrated this problem using topology analysis on $G$~\cite{okumura2023lacam2,matsui2024investigation,zhou2025loosely}, sometimes called the \emph{swap} technique.
Meanwhile, this study aims to improve solution quality in broader situations, not limited to narrow corridors.
Similar to the guidance, the swap can co-enhance PIBT with the proposed techniques.

Recent work has discovered direct preference optimisation using data-driven approaches, in particular imitation learning from expert (near-)optimal algorithms~\cite{veerapaneni2024improving,jiang2024deploying}.
Although this direction is attractive, it requires a significant amount of offline preparation.
Another concern with data-driven artefacts is the notably slow inference~\cite{skrynnik2024learn} compared to the heuristic search-based preference of \cref{eq:preference}, which is a bottleneck when incorporating PIBT into a search scheme.
Rather, Monte-Carlo sampling over PIBT, i.e. running vanilla PIBT several times with random tiebreaking and retrieving the best one as a successor node, is fast and effective with the search~\cite{okumura2024lacam3}.

\section{Hindrance}
A \hindrance term aims to achieve smarter yet effective dodge behaviour when another agent is about to move to the currently occupying location.
In \cref{fig:hindrance}(a,c), the lower priority agent must avoid the higher priority agent by either going right or down.
If it goes right (\cref{fig:hindrance}b), the low-priority agent still has to avoid the high-priority agent at the next timestep, resulting in a minimum of five steps to reach its goal.
Meanwhile, if it goes down (\cref{fig:hindrance}d), this agent can reach the goal faster, i.e., in three steps.
The preference construction with \cref{eq:preference} cannot distinguish these two situations.

{
\begin{figure}[t!]
\centering
\includegraphics[height=0.7\linewidth]{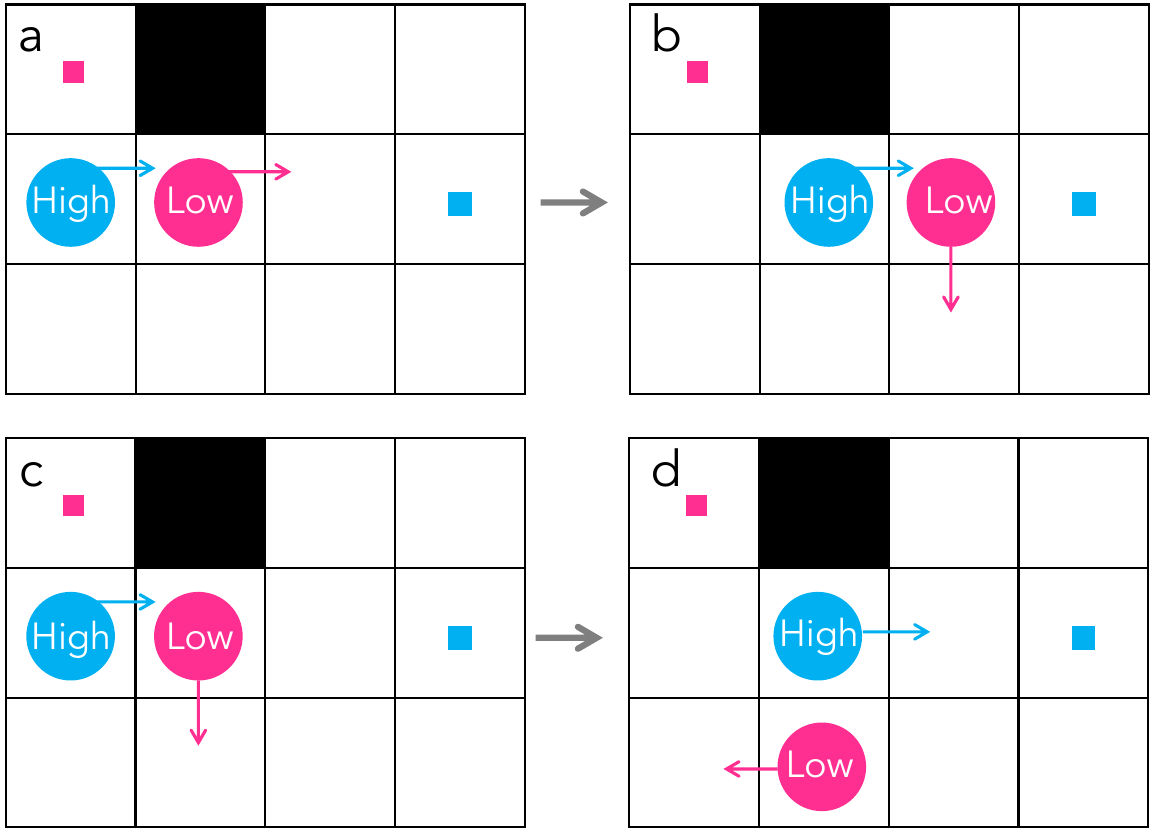}
\caption{
Motivating example for calculating the \hindrance term.
Goals for the agents are marked with coloured boxes.
}
\label{fig:hindrance}
\end{figure}
}

{
  \begin{algorithm}[t!]
    \caption{Hindrance calculation}
    \label{algo:hindrance}
    \begin{algorithmic}[1]
      \small
      \Input{configuration $\Q$, agent $i \in A$, location $u \in V$}
      \State $\mathit{hindrance} = 0$
      \For{$j \in \{ k \in A \mid Q[k] \in \neigh(\Q[i]) \}$}
      \If{$u \neq Q[j] \land \dist(u, g_j) < \dist(\Q[i], g_j)$}
      \label{algo:hindrance:check}
      \State $\mathit{hindrance} \leftarrow \mathit{hindrance} + 1$
      \EndIf
      \EndFor\\
      \Return $\mathit{hindrance}$
    \end{algorithmic}
  \end{algorithm}
}

Intelligent tiebreaking in preference construction is possible by estimating how such an avoidance behaviour by the low-priority agent might still disrupt the progress of the high-priority agent in the next timestep.
Specifically, this estimate, called \hindrance, is computed by examining only the distance relations between several vertices, as shown in \cref{algo:hindrance}.
For a potential action $u \in V$ for an agent $i \in A$, the low-priority agent in \cref{fig:hindrance}, \cref{algo:hindrance} checks whether $u$ is heading towards a goal for a neighbouring agent $j \in A$, i.e., high-priority agent.
Such an action is summed up as \hindrance, enabling PIBT to distinguish between the cases of \cref{fig:hindrance}a and \cref{fig:hindrance}c.
Note that the first condition in \cref{algo:hindrance:check} does not penalise non-dodging behaviour, e.g. going left for the low-priority agent in \cref{fig:hindrance}a.

\paragraph{Time Complexity.}
\Cref{algo:hindrance} is a computationally lightweight procedure.
Under the assumption that $\dist(\cdot, g_j)$ is constant,%
\footnote{
Most MAPF implementations first compute the distance table using backward Dijkstra, and then look up this table to retrieve the distance information in constant time during the search.
}
\cref{algo:hindrance} has mere $O(\Delta)$ time complexity, where $\Delta$ is the maximum degree of a graph $G$, e.g., four in four-connected gridworld, on which MAPF methods are typically tested.
Consequently, the total overhead for the PIBT step remains $O(|A|\cdot\Delta^2)$ because each agent computes \hindrance once for each action.

\section{Regret Learning}

While \hindrance improves how low-priority agents avoid high-priority ones, this section aims to improve how high-priority agents break ties among multiple actions.
We calculate how each action affects \regret of other agents, the cost gap between the best action and the action actually taken, and use it for tiebreaking.
This term is expected to be effective in highly congested situations where the chain of priority inheritance is often triggered within PIBT.

{
\begin{figure}[t!]
\centering
\includegraphics[height=0.8\linewidth]{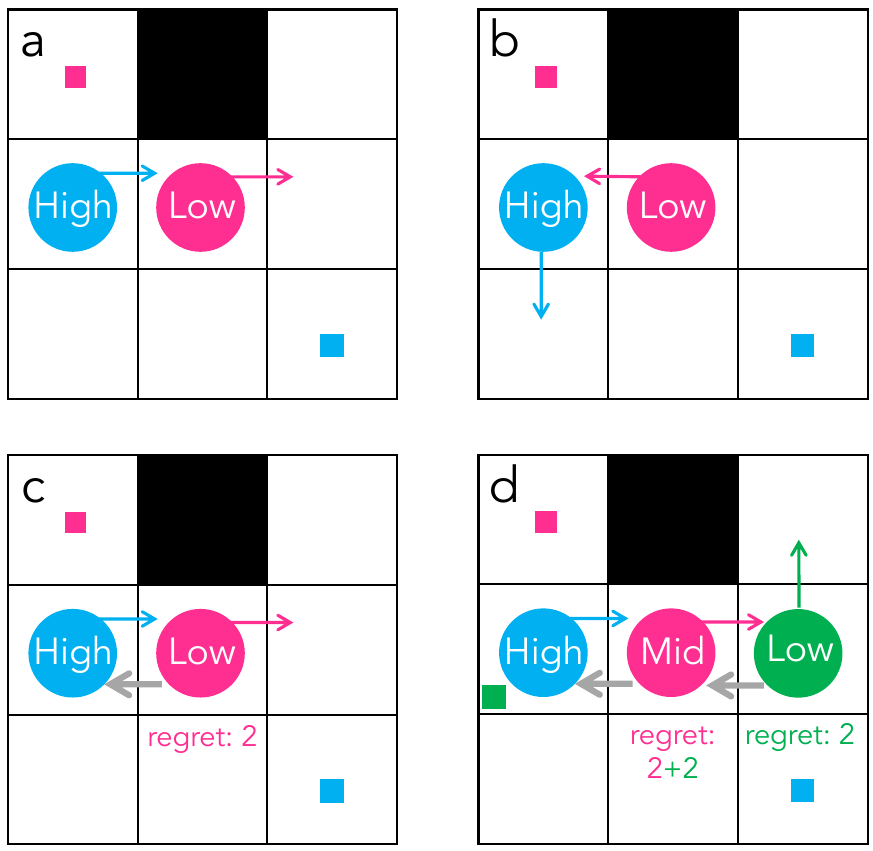}
\caption{
Motivating example for calculating the \regret term.
Grey arrows represent backtracking in PIBT.
}
\label{fig:regret}
\end{figure}
}

{
  \renewcommand{\P}{\m{\mathcal{R}}}
  \begin{algorithm}[t!]
    \caption{PIBT with regret propagation}
    \label{algo:regret}
    \begin{algorithmic}[1]
      \small
      \Param{$m \in \mathbb{N}_{>0}, 0 \leq w \leq 1$}
      \For{$i \in A$}
      \Comment{initialise regret table}
      \label{algo:regret:init-table}
      \For{$v \in \neigh\left(\Q\from[i]\right) \cup \left\{\Q\from[i]\right\}$}
      \State $\P[i, v] \leftarrow 0$
      \EndFor
      \EndFor
      \label{algo:regret:end-init}
      \For{$1, 2, \ldots, m$}
      \label{algo:regret:learning-start}
      \Comment{learning regret}
      \State initialise  $\Q\to$
      \State \same{\textbf{for}~$i \in A$~\textbf{do};~
             \textbf{if}~{$\Q\to[i] = \bot$}~\textbf{then}~{$\PIBT(i)$}}
      \EndFor
      \label{algo:regret:learning-end}
      \State \same{\Return $\Q\to$}
      \item[]
      \ProcedureSame{\PIBT}{$i$}
      \State \same{$C \leftarrow \neigh\left(\Q\from[i]\right) \cup \left\{\Q\from[i]\right\}$}
      \State sort $C$ with $\P[i, \cdot]$ as tiebreaker
      \ForSame{$v \in C$}
      \IfSingleSame{collisions in $\Q\to$ supposing $\Q\to[i]=v$}{\Continue}
      \State \same{$\Q\to[i] \leftarrow v$;} $\regret \leftarrow 0$
      \IfSame{$\exists j \in A~\text{s.t.}~j \neq i \land \Q\from[j]=v \land \Q\to[j]=\bot$}
      \State $\mathit{validity},  \regret \leftarrow \PIBT(j)$
      \State $\P[i, v] \leftarrow (1 - w) * \P[i, v] + w * \mathit{regret}$
      \label{algo:regret:update-table}
      \IfSingleSame{$\mathit{validity} = \invalid$}{\Continue}
      \EndIf
      \State $\regret_i \leftarrow \dist(v, g_i) - \min_{u \in C} \dist(u, g_i)$
      \State \same{\Return~\valid,} $\regret + \regret_i$
      \label{algo:regret:backtracking1}
      \EndFor
      \State $\mathit{regret}_i \leftarrow \dist(\Q\from[i], g_i) - \min_{u \in C} \dist(u, g_i)$
      \State \same{$\Q\to[i] \leftarrow \Q\from[i]$;~\Return~\invalid,} $\mathit{regret}_i$
      \label{algo:regret:backtracking2}
      \EndProcedure
    \end{algorithmic}
  \end{algorithm}
}

\Cref{fig:regret}(a,b) describes example situations to see the importance of high-priority agent action selection.
From the perspective of the high-priority agent, going either right or down brings the agent closer to its goal, while the low-priority agent cannot take the shortest path if the right action is chosen.
Instead, the low-priority agent experiences \regret, computed by $\dist(v, g_i) - \min_{u \in C} \dist(u, g_i)$, where $v$ is the selected action and $C$ is the set of candidate actions.
With PIBT, it is easy to implement so that the high-priority agent can know \regret of others through backtracking (\cref{fig:regret}c).
The same scheme applies when priority inheritance occurs recursively (\cref{fig:regret}d).
Therefore, when we run PIBT again, based on this information, the high-priority agent can choose the down action to avoid experiencing others with \regret.

This regret learning is a non-stationary process, as in each PIBT run, each agent changes its action based on the current \regret estimate, and then \regret might be different even if one took the same action.
Our simple solution to this is running PIBT several times, and over the iterations, we gradually update the estimates while incorporating the past estimates as a weighted average.

\Cref{algo:regret} embodies the concept above, while greying out the same lines from the original PIBT (\cref{algo:pibt}).
Lines~\ref{algo:regret:init-table}--\ref{algo:regret:end-init} first initialise the regret table $\mathcal{R}$, which records \regret from others when an agent $i \in $ takes action $v \in V$.
The regret table is learned through several PIBT runs (Lines~\ref{algo:regret:learning-start}--\ref{algo:regret:learning-end}), and the final run is used as output.
The subprocedure requires minor modifications to include \regret in backtracking, in addition to the validity of priority inheritance (\cref{algo:regret:backtracking1,algo:regret:backtracking2}).
For each priority inheritance call, the regret table is updated with a weighted sum of the previous and new values (\cref{algo:regret:update-table}).

\paragraph{Time Complexity.}
Regret learning is simply running the original PIBT several times, typically three times is sufficient to improve the solution quality.
This does not significantly affect the speed advantage of PIBT, which runs $O(|A|\cdot\Delta)$ in its minimal form, empirically sub-millisecond procedure even with thousands of agents.
Besides, regret learning is smarter than simply using the presence of another agent, which is used in the original PIBT paper~\cite{okumura2022priority}, because that ad-hoc rule does not necessarily involve regret.
Monte-Carlo configuration generation~\cite{okumura2024lacam3} similarly runs PIBT several times and retrieves the best one according to the heuristic, but there is no guidance on sampling sequences to improve its quality over trials.
Consequently, regret learning is more efficient in terms of sampling efficiency.
We will empirically observe these claims in the next section.

{
\newcommand{\entry}[1]{
  \begin{minipage}{0.23\linewidth}
    \centering
    \hspace{0.1\linewidth}
    \begin{minipage}{0.23\linewidth}
      \includegraphics[height=1.0\linewidth]{fig/raw/map/#1}
    \end{minipage}
    \begin{minipage}{0.4\linewidth}
      \hspace{0.2em}
      {\small\mapname{#1}}
    \end{minipage}
    \includegraphics[width=1\linewidth,trim={0.5cm 0 1cm 1cm},clip]{fig/raw/#1-oneshot-SoCS2025}
  \end{minipage}
}

\begin{figure*}[t!]
  \centering
  \entry{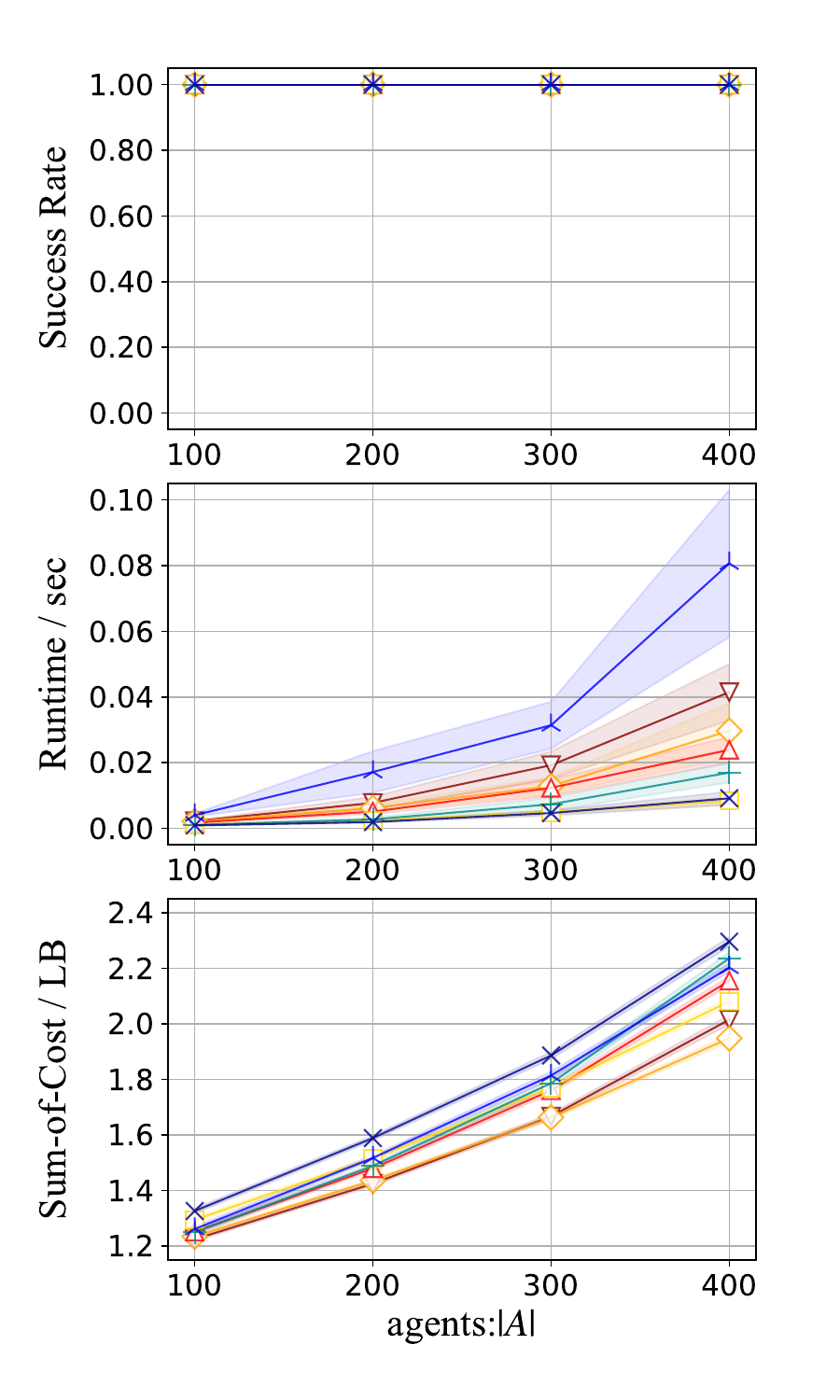}
  \entry{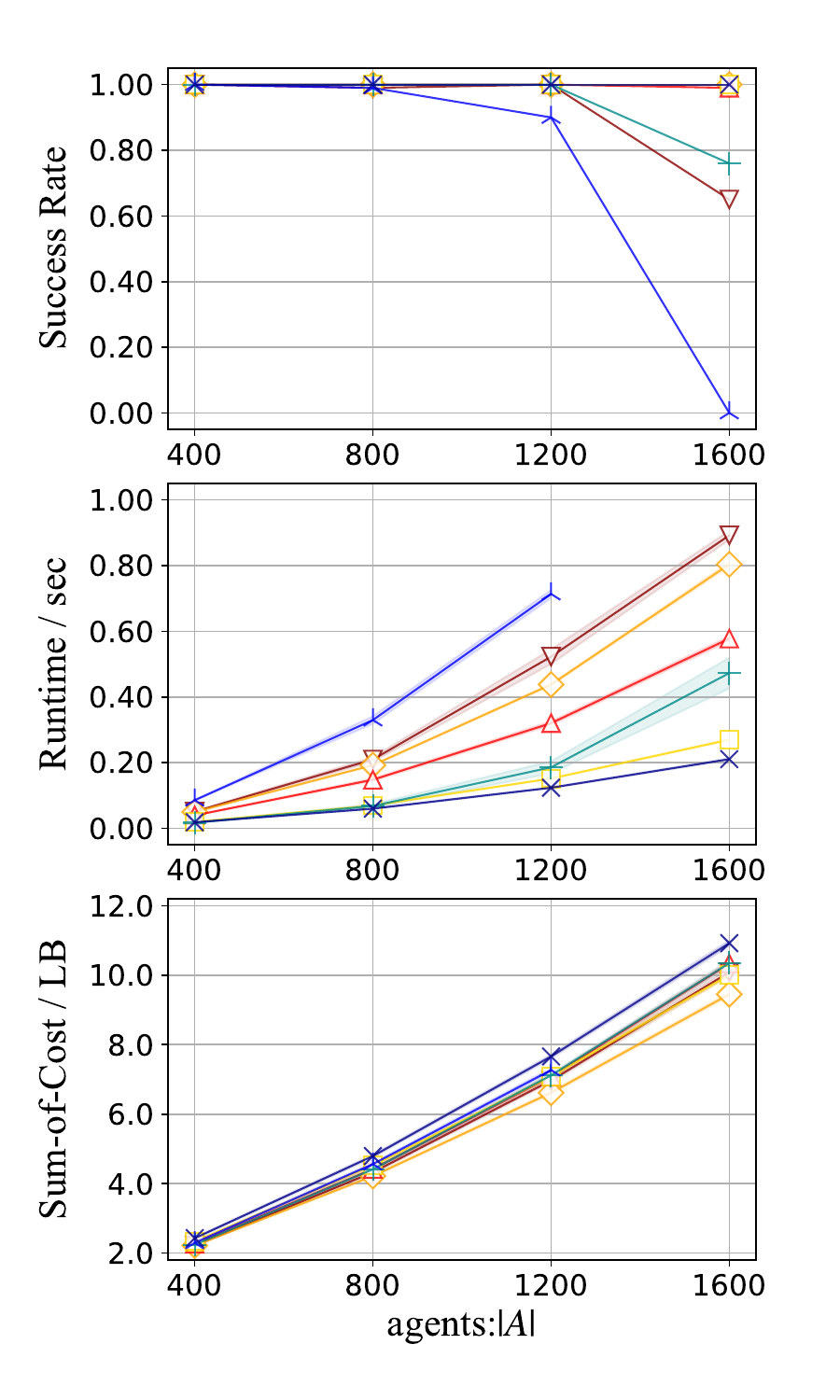}
  \entry{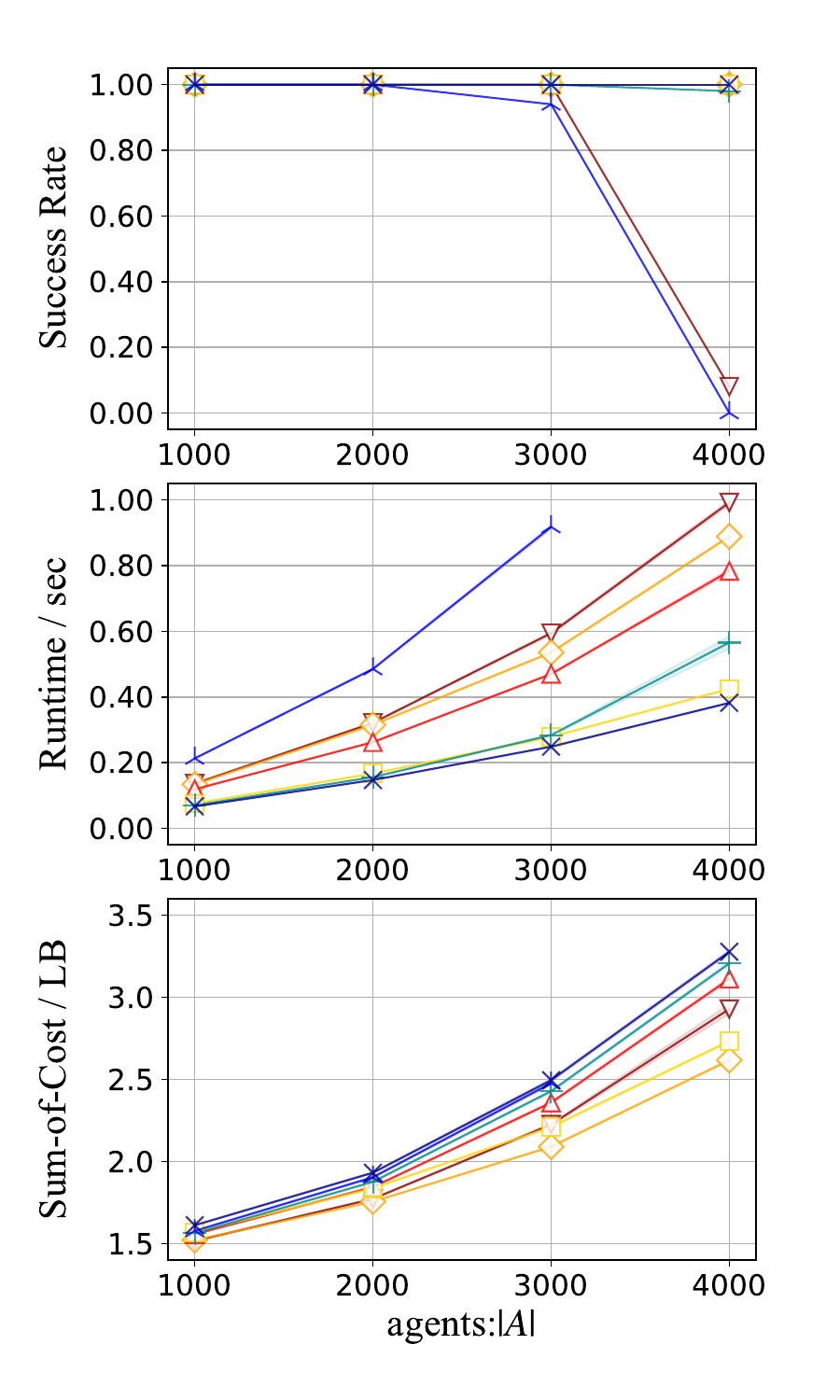}
  \entry{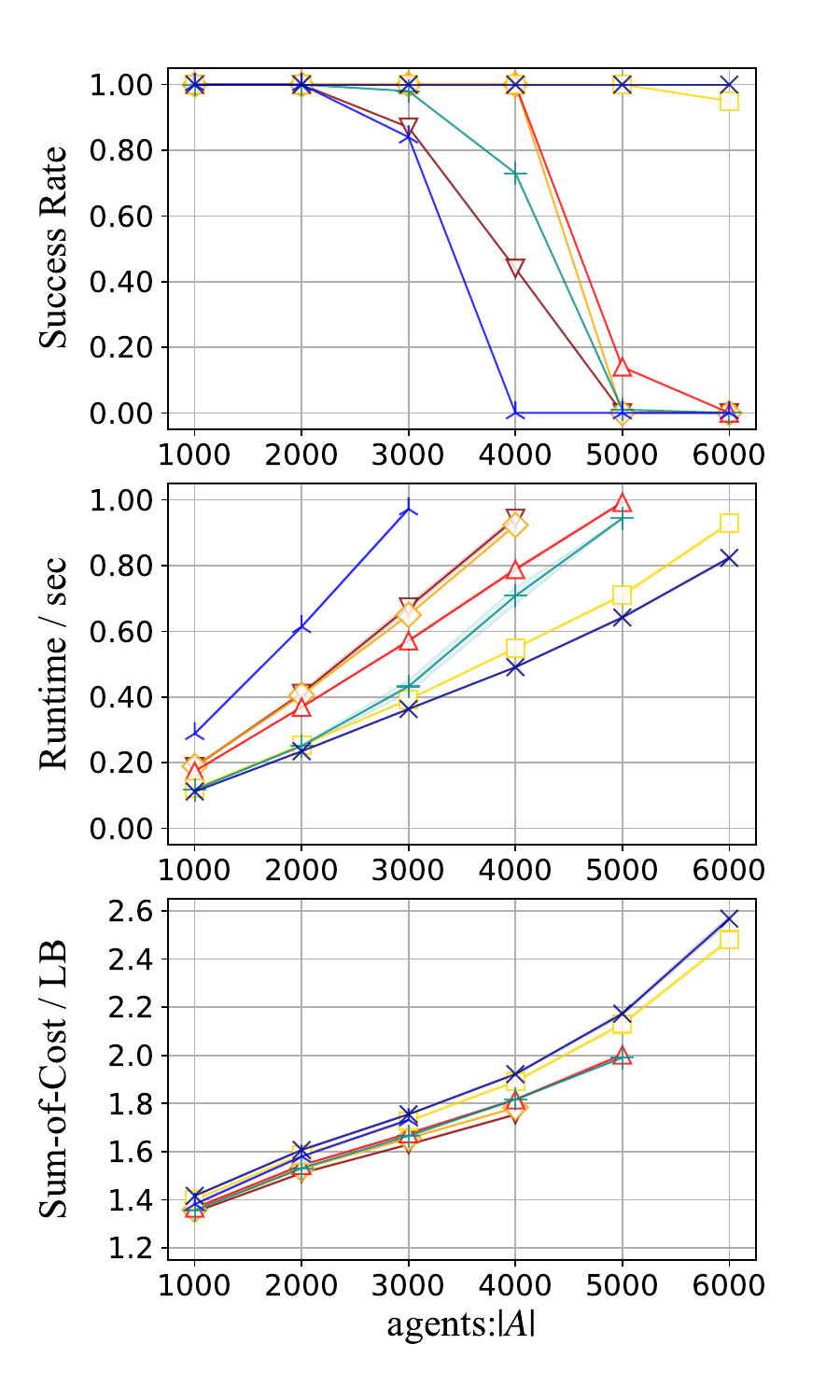}
  \includegraphics[width=1\linewidth]{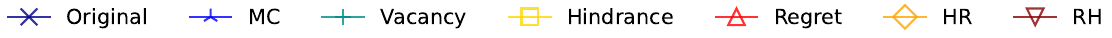}
  \caption{
    Results for one-shot MAPF.
    The success rate of planning within \SI{1}{\second} (top), average runtime (middle) and SoC normalised by lower bound ($1.0$ is minimum; bottom) for successful cases are shown.
  }
  \label{fig:oneshot-mapf}
\end{figure*}
}

{
\newcommand{\entry}[1]{
  \begin{minipage}{0.23\linewidth}
    \centering
    {\small\mapname{#1}}\\
    \includegraphics[width=1\linewidth,trim={0.5cm 0 1cm 0.5cm},clip]{fig/raw/#1-lifelong-SoCS2025}
  \end{minipage}
}
\begin{figure*}[t!]
  \centering
  \entry{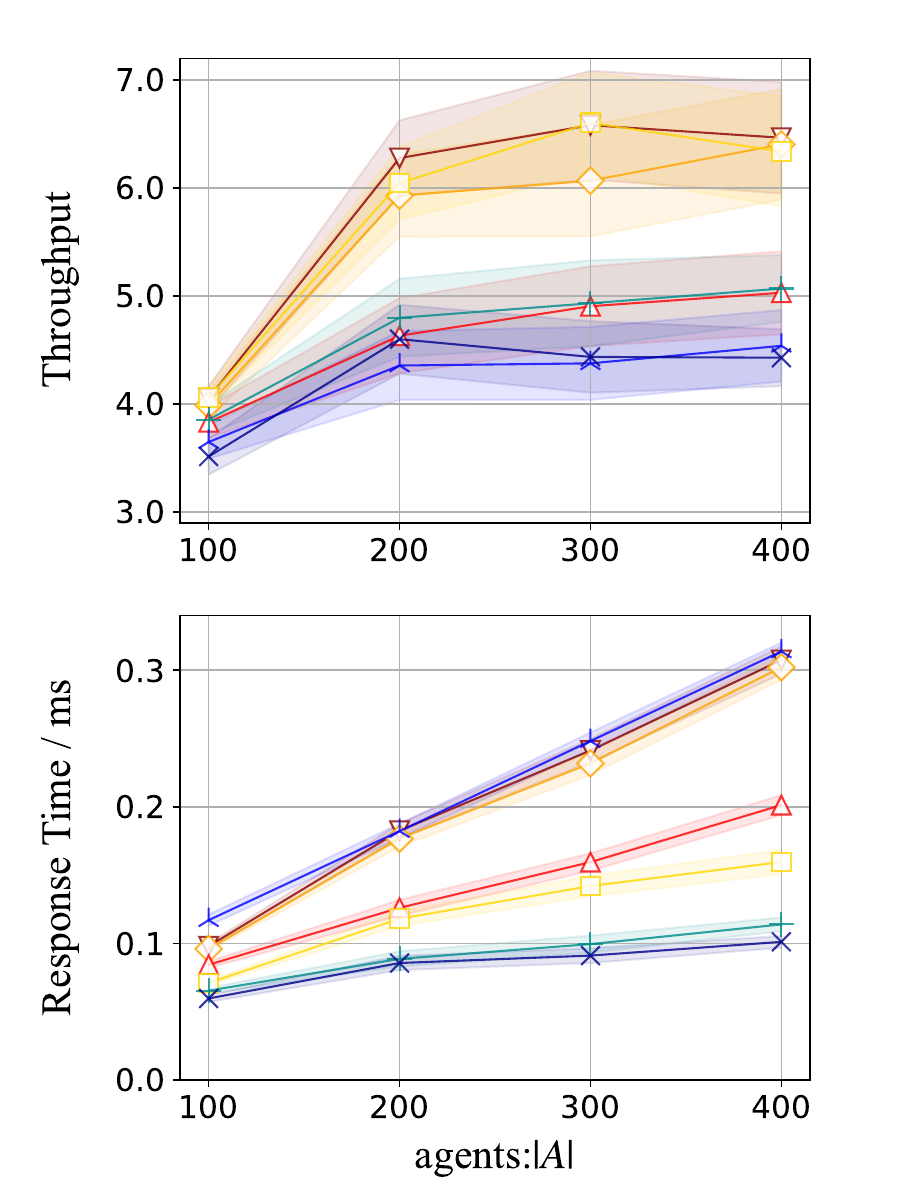}
  \entry{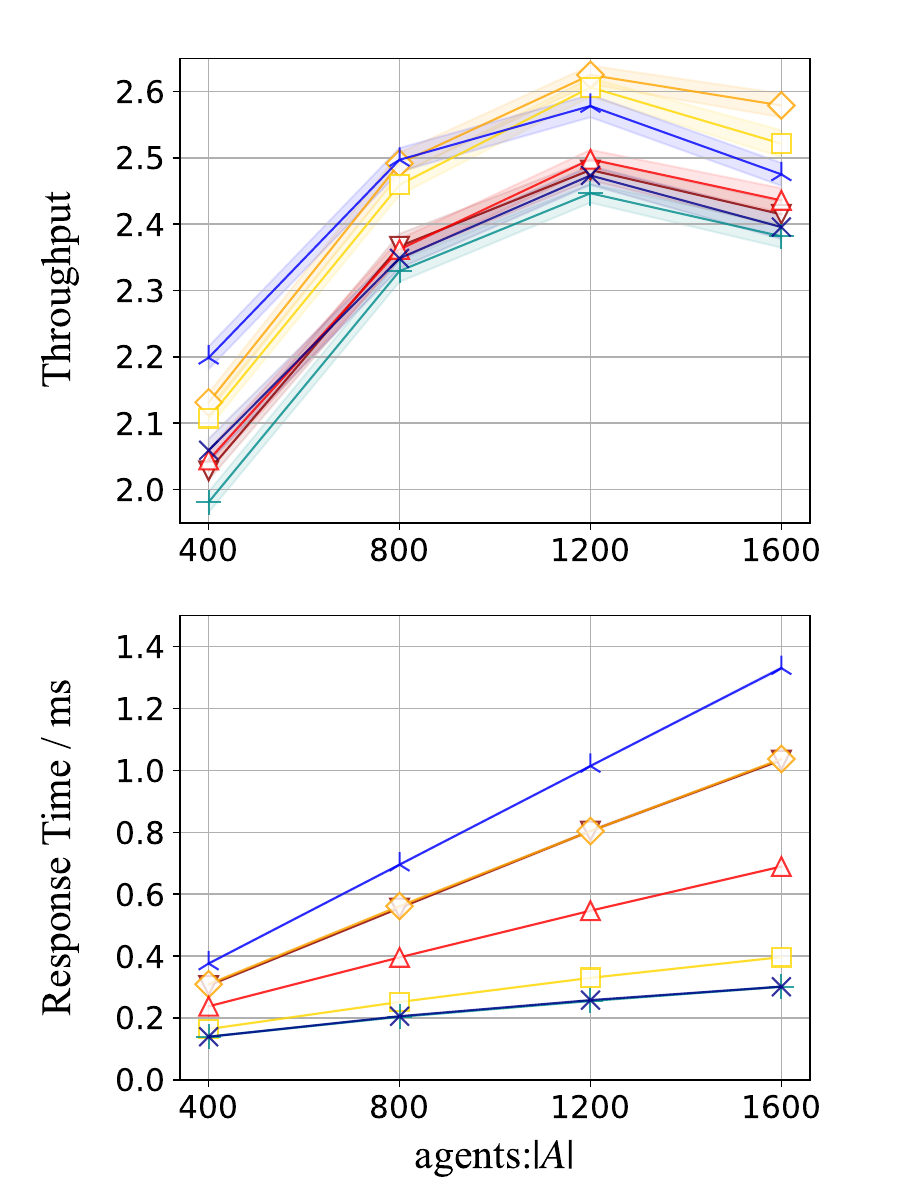}
  \entry{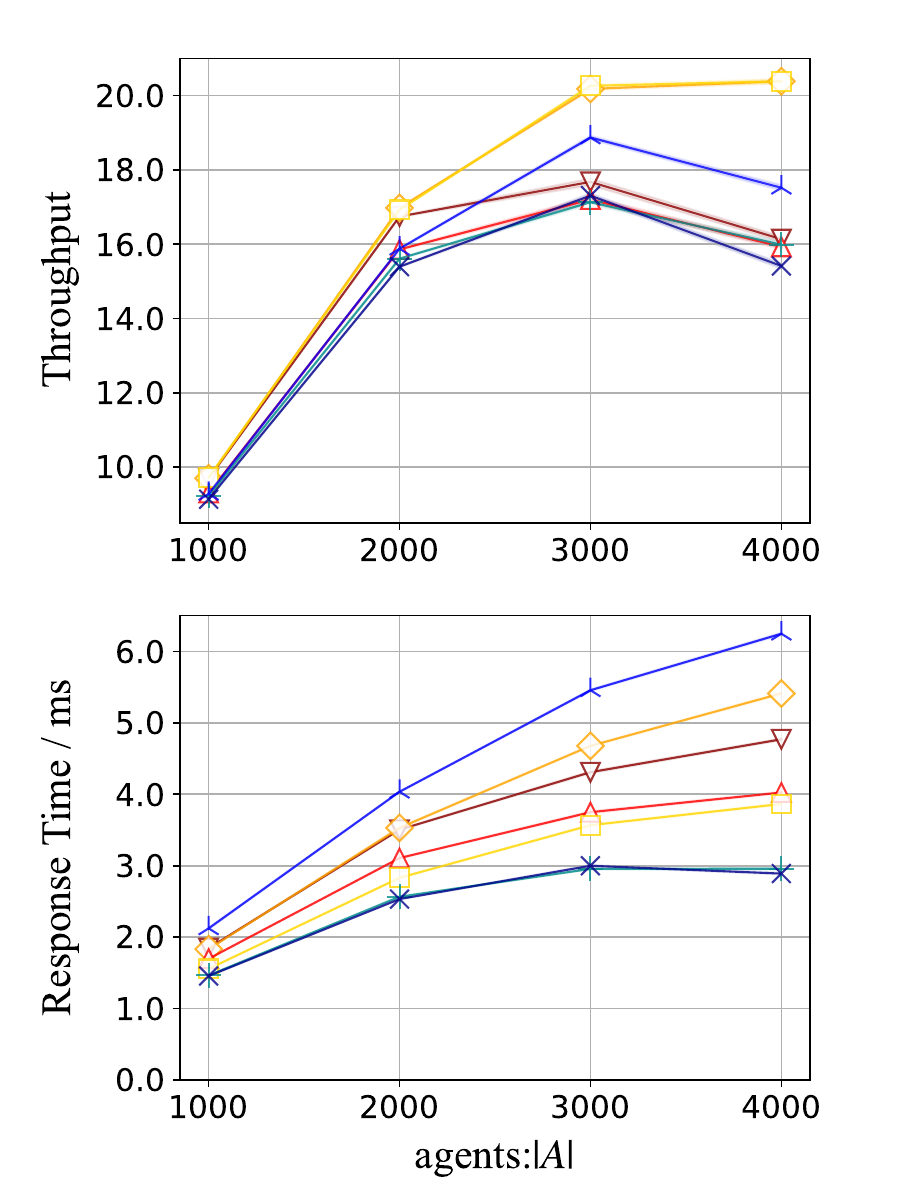}
  \entry{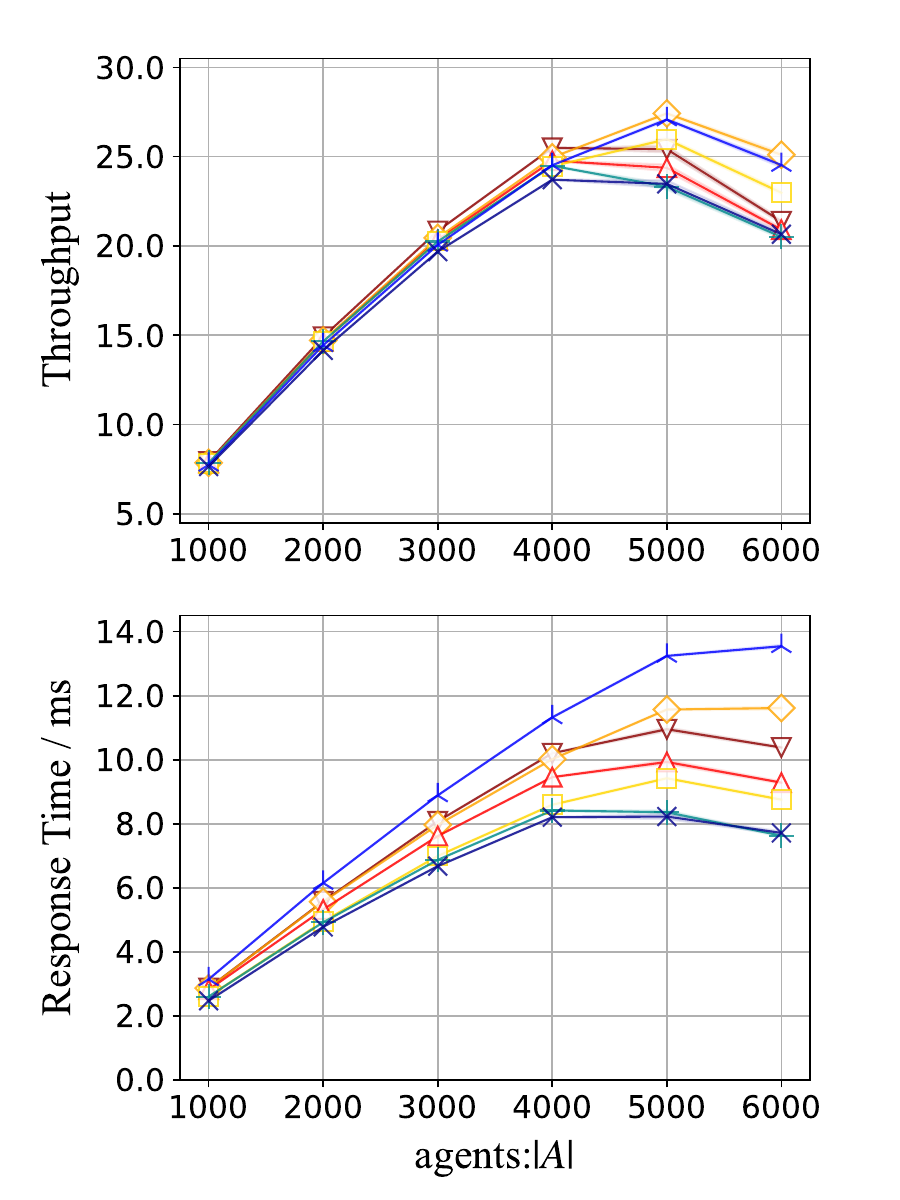}
  \includegraphics[width=1\linewidth]{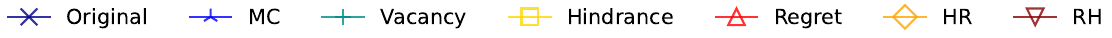}
  \caption{
    Results for lifelong MAPF.
    Average throughput (top) and response time (bottom) are shown.
  }
  \label{fig:lifelong-mapf}
\end{figure*}
}

\section{Experiments}
Through both one-shot and lifelong MAPF problems, we empirically evaluate several tiebreaking strategies on PIBT, including the proposed techniques, as listed below.
\begin{itemize}
\item We refer to \strategy{Original} as preference construction using the vanilla method, i.e. ascending order with \cref{eq:preference}.
\item \strategy{MC}, which stands for Monte Carlo sampling, is a strategy introduced in the latest LaCAM implementation~\cite{okumura2024lacam3}.
  From a given configuration $\Q$, it creates a batch $\mathcal{B}$ of $k$ configurations using PIBT and \cref{eq:preference} but with different random seeds.
  \strategy {MC} then selects the cost-minimising configuration as $\argmin_{\Q' \in \mathcal{B}}(\funcname{g}(\Q, \Q') + \funcname{h}(\Q'))$, where \funcname{g} is a transition cost $|\{ i \in A \mid \neg (\Q[i] = \Q'[i] = g_i) \}|$ and \funcname{h} is a heuristic $\sum_{i \in A}\dist(Q'[i], g_i)$.
  The experiments use $k=10$ following the original work.
\item \strategy{Vacancy} follows the PIBT paper~\cite{okumura2022priority}, which prioritises a vacant location if available.
  Formally, the preference is constructed with
  $\left\langle \dist(v, g_i), \mathrm{Ind}\big[\exists j \in A, v = \Q[j]\big], \epsilon \right\rangle$,
  where $\mathrm{Ind}[\cdot] = 1$ if the condition is true; zero otherwise.
\item \strategy{Hindrance} uses $\langle \dist, \hindrance, \epsilon \rangle$ for the preference construction.
\item \strategy{Regret} uses $\langle \dist, \regret, \epsilon \rangle$.
  Unless explicitly mentioned, the parameters are set to $m=3$ and $w=0.9$.
  These parameters have been adjusted following pilot studies to give reasonable results over a wide range of situations while minimising $m$.
\item \strategy{HR} uses $\left\langle \dist, \hindrance, \regret, \epsilon \right\rangle$.
\item \strategy{RH} uses $\left\langle \dist, \regret, \hindrance, \epsilon \right\rangle$.
\end{itemize}
The prioritisation scheme in PIBT follows the original paper.

\paragraph{Benchmarks.}
Experiments were conducted on a laptop equipped with an M3 Pro Apple Silicon chip and \SI{18}{\giga\byte} of RAM, using the following four-connected grid maps:
\begin{itemize}
\item \mapname{Random} (\mapname{random-32-32-10}) is a $32\times32$ grid map with 10\% obstacles and $|V|=922$.
\item \mapname{Room} (\mapname{room-64-64-8}) sizes $64\times 64$ and $|V|=3,232$.
\item \mapname{Warehouse} (\mapname{warehouse-10-20-10-2-2}) represents a typical warehouse layout with $170\times 84$ and $|V|=9,766$.
\item \mapname{Sortation}, with $200\times 140$ and $|V|=21,920$, has single cell obstacles evenly spaced for every two columns/rows.
\end{itemize}
The first three are from the MAPF benchmark~\cite{stern2019def}, while the last is from~\cite{chen2024traffic}.
Their illustration is available in \cref{fig:oneshot-mapf}.
The values reported were generated from 100 test cases, prepared for each map and each number of agents, with the start and goal positions randomised.
The code is written in C++ and is available at \url{https://github.com/HirokiNagai-39/pibt-tiebreaking}.

\subsection{One-shot MAPF}
In the context of one-shot MAPF, there is a popular search algorithm called LaCAM~\cite{okumura2023lacam} on top of PIBT that outperforms naively running PIBT.
We thus evaluate tiebreaking strategies through LaCAM.
\emph{Our implementation is based on the vanilla LaCAM from \cite{okumura2023lacam}, which does not use the advanced techniques introduced later in~\cite{okumura2023lacam2,okumura2024lacam3} to isolate the effect of tiebreaking.}
Note that the original LaCAM paper uses \strategy{Original} as its tiebreaking strategy.

We assess the tiebreaking effect with
\emph{(i)}~planning success rate within $\SI{1}{\second}$ of finding a feasible solution,
\emph{(ii)}~wall clock time for finding a solution, and
\emph{(iii)}~solution cost represented by SoC (sum-of-costs).
The SoC value shown is normalised by the sum of the shortest path lengths for all agents between their starts and goals, i.e. the trivial lower bound, so the minimum is one.
The \SI{1}{\second} timeout reflects the real-world demands of real-time planning in logistics systems, and is also used in the MAPF's League of Robot Runners competition~\cite{chan2024league}.

\Cref{fig:oneshot-mapf} summarises the results, showing that overall, both \hindrance and \regret tiebreakers improve the planning ability of LaCAM in dense and challenging MAPF scenarios involving several hundreds to thousands of agents, with little computational overhead.
In other words, these tiebreaking strategies in PIBT serve to better guide the search towards the goal configuration with less redundant agent movements, compared to existing ones such as \strategy{MC} and \strategy{Vacancy}.
The joint use of \hindrance and \regret further enhances the performance of LaCAM.
Notably, \strategy{HR} achieves approximately 20\% cost reduction from \strategy{Original} in \mapname{warehouse} with $4,000$ agents.

The \hindrance term is particularly effective with minimal engineering.
At the extreme end, \strategy{Hindrance} has a success rate of over 90$\%$ on \mapname{sortation} with $6,000$ agents as like \strategy{Original}, even with a strict time limit, still improving LaCAM's solution quality in \emph{all} scenarios.
Meanwhile, the quality of the resulting solution varies from map to map.
In \mapname{random} and \mapname{warehouse}, LaCAM results in smaller SoC solutions with \strategy{Hindrance}, while \strategy{Regret} performs better in \mapname{sortation} in terms of the solution quality.
We presume that this is because \mapname{sortation} has a structure that \regret is effectively propagated due to its regular obstacle placements.
These map-dependent results have a direct impact on the order in which these terms should be used, as we can see with \strategy{HR} and \strategy{RH}.

The \regret term requires PIBT to be run several times, which does indeed affect the runtime and causes some timeout attempts in instances with massive agents.
However, it should be noted that the absolute runtime difference remains within hundreds of milliseconds even with thousands of agents in the cases tested.
Furthermore, \strategy{Regret} outperforms \strategy{MC} with a similar concept, which also requires multiple PIBT runs.
This shows that \regret captures complex interactions with agents better than `blind' trials of \strategy{MC}, due to its improved backtracking process.

\paragraph{Effect on Hyperparameters in Regret Learning.}
The construction of the \regret term requires two hyperparameters: the number of learning iterations $m$ and the weight $w$, to sum up the regret values from different iterations.
\Cref{tab:ablation} examines how these parameters affect the performance of \strategy{Regret} using two scenarios.
\strategy{Regret} has $m$ times the overhead of vanilla PIBT according to the time complexity analysis.
The runtime results roughly follow this observation, with slight deviations due to LaCAM's search process.
The SoC metric generally improves as $m$ increases with more accurate regret learning, eventually reaching saturation.
In contrast, the weight $w$ is not particularly dominant.

{
\newcommand{\e}[1]{{\scriptsize$\pm$#1}}

\begin{table}[tp!]
\centering
\mapname{random}, $|A|=400$
\begin{tabular}{rrrrr}
\toprule
$m$ && $w=0.5$ & $w=0.9$ & $w=0.95$
\\\midrule
\multirow{2}{*}{3} & time & 0.029\e{0.007} & 0.024\e{0.004} & 0.029\e{0.008}
\\
& SoC & 2.161\e{0.017} & 2.155\e{0.017} & 2.157\e{0.016}
\\
\midrule
\multirow{2}{*}{10} & time & 0.136\e{0.037} & 0.110\e{0.025} & 0.112\e{0.019}
\\
& SoC & 2.118\e{0.024} & 2.110\e{0.022} & \textbf{2.101}\e{0.022}
\\
\midrule
\multirow{2}{*}{20} & time & 0.267\e{0.054} & 0.231\e{0.041} & 0.212\e{0.044}
\\
& SoC & \textbf{2.101}\e{0.021} &  2.114\e{0.028} & \textbf{2.101}\e{0.023}
\\
\bottomrule
\end{tabular}\medskip\\
\mapname{warehouse}, $|A|=4,000$
\begin{tabular}{rrrrr}
\toprule
$m$ && $w=0.5$ & $w=0.9$ & $w=0.95$
\\\midrule
\multirow{2}{*}{3} & time & 0.788\e{0.007} & 0.780\e{0.006} & 0.792\e{0.007}
\\
& SoC & 3.122\e{0.010} & 3.112\e{0.010} & 3.112\e{0.010}
\\
\midrule
\multirow{2}{*}{10} & time & 2.876\e{0.084} & 2.871\e{0.115} & 2.817\e{0.107}
\\
& SoC & 2.929\e{0.009} & 2.918\e{0.010} & 2.913\e{0.009}
\\
\midrule
\multirow{2}{*}{20} & time & 6.888\e{0.299} & 6.398\e{0.291} & 6.369\e{0.312}
\\
& SoC & \textbf{2.897}\e{0.011} & 2.904\e{0.010} & 2.901\e{0.010}
\\
\bottomrule
\end{tabular}
\caption{
  Effect on hyperparameters for \strategy{Regret} over 100 instances, resulting in all successful.
  The time unit is seconds.
  \SI{10}{\second} timeout is used.
}
\label{tab:ablation}
\end{table}
}

{
\begin{figure*}[t!]
\centering
\includegraphics[width=1\linewidth,clip]{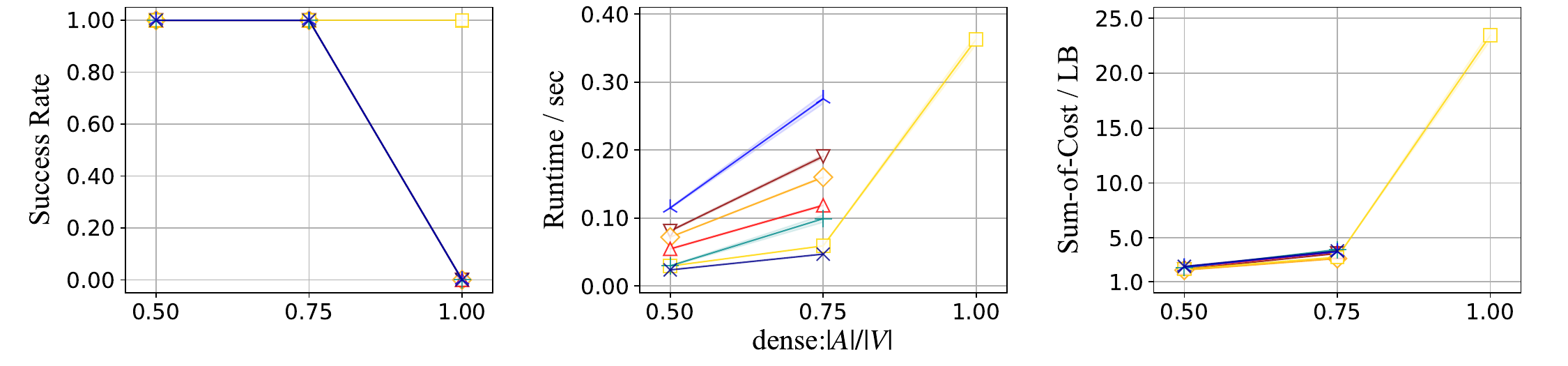}
\includegraphics[width=1\linewidth]{fig/raw/legend2}
\caption{
One-shot MAPF on \mapname{empty} in extremely dense situations.
}
\label{fig:ex-dense}
\end{figure*}
}

\subsection{Lifelong MAPF}
Next, we evaluate tiebreaking strategies of PIBT through a popular variant of MAPF such that once an agent reaches the goal, a new goal is randomly assigned.
There are two metrics of interest here:
\emph{(i)}~throughput, the number of goals reached for each timestep, averaged over the operation period, $1,000$ in our case, and
\emph{(ii)}~response time required to generate a plan for each timestep.

\Cref{fig:lifelong-mapf} summarises the results.
In all experimental conditions, the difference in response time between the different tiebreaking strategies is almost negligible ($\leq$\SI{10}{\milli\second}) in absolute terms, even with thousands of agents.
Meanwhile, the throughput is significantly improved with \hindrance in all the maps tested;
for example, \strategy{Hindrance}, \strategy{HR}, and \strategy{RH} all achieve $\geq 40\%$ throughput improvements with 400 agents in \mapname{random}.
The \regret term also has steady improvements over \strategy{Original}, but its effect is subtle compared to \hindrance.
We do not have solid interpretations for this;
perhaps a one-step optimisation of the regret values would not correlate strongly with the throughput improvement.

\subsection{Extremely Dense Situations}
We further investigate the capability of newly developed tiebreaking strategies in extremely dense one-shot MAPF instances.
Specifically, using the \mapname{empty} map ($48\times 48$; $|V|=2,304$) taken from~\cite{stern2019def}, we prepare instances with $|A|/|V|=\{0.5, 0.75, 1.0\}$, i.e., $1,152$, $1,728$, and $2,304$ agents, respectively.
This time the time limit is set to \SI{10}{\second}, taking into account the difficulty.

{
  \setlength{\tabcolsep}{5pt}
  \newcommand{\e}[1]{{\scriptsize$\pm$#1}}
  \begin{table}[t!]
    \centering
    \mapname{empty}, $|A| = |V| = 2,304$
    \begin{tabular}{rrrrr}
      \toprule
      $m$ && $w=0.5$ & $w=0.9$ & $w=0.95$
      \\
      \midrule
      \multirow{2}{*}{20} & time & 3.837\e{0.067} & 3.657\e{0.057} & 3.649\e{0.047}
      \\
      & SoC & 25.140\e{0.321} & 24.136\e{0.276} & \textbf{24.129}\e{0.248}
      \\

      \midrule
      \multirow{2}{*}{30} & time & 5.737\e{0.066} & 5.614\e{0.008} & 5.460\e{0.067}
      \\
      & SoC & 26.234\e{0.196} & 25.145\e{0.195} & 24.874\e{0.201}
      \\
      \bottomrule
    \end{tabular}
    \caption{
      Performance of \strategy{Regret} with the extremely dense scenario where $|A|/|V|=1$.
    }
    \label{tab:ex-dense-regret}
  \end{table}
}

\Cref{fig:ex-dense} presents the results.
Remarkably, LaCAM with \strategy{Hindrance} was able to solve $|A|/|V|=1$ instances with a success rate of $100\%$, within \SI{1}{\second}, while \strategy{Original} completely fails in the same setting.
This observation provides further evidence that \hindrance can serve as a stronger guide for the search than doing nothing to break a tie.

The failures in \strategy{HR} with high density imply that adding \regret leads the search in the wrong direction.
This is actually caused by immature regret learning with insufficient PIBT iteration specified by the parameter $m$, currently set to three.
\Cref{tab:ex-dense-regret} presents this evidence, showing that increasing $m$ allows LaCAM to solve extremely dense instances with \strategy{Regret} only.
With $m=\{20, 30\}$, \strategy{Regret} results in a success rate of $100\%$, while lower $m$ (e.g., 10) could not solve any instances regardless of $w$.
Their solution quality is comparable to \strategy{Hindrance}, indicating that \regret is a viable option in itself.

\subsection{Discussions}
Overall, adding \hindrance consistently improves the performance of PIBT (\strategy{Original}) and outperforms the other strategies in terms of solution cost and computational overhead.
The \regret term is generally effective, but not as powerful as \hindrance;
nevertheless, in one-shot MAPF, it improves LaCAM compared to \strategy{MC}, which shares the concept of running PIBT several times.
Regret learning is understood as a general scheme for \strategy{Vacancy}, and thus has a similar effect on planning.
Meanwhile, it could have the potential to reflect complicated agent interactions thanks to regret propagation, as seen in \cref{tab:ex-dense-regret}.
\strategy{MC} can enhance the vanilla PIBT and is especially promising in lifelong MAPF, but not stable in one-shot MAPF.
We note that adding the so-called swap technique~\cite{okumura2023lacam2} may clear out this instability, but still, \regret could be a better alternative given the same amount of time budget.

Considering these observations, our suggestion is to use \cref{eq:hindrance-regret} as PIBT tiebreaking.
In fact, \strategy{HR} consistently achieves superior solution quality among tested strategies in both one-shot and lifelong MAPF scenarios, with a slight runtime addition to \strategy{Original}.

\section{Conclusion}
This paper examined the technical details of PIBT that underpins modern large-scale MAPF studies.
Our focus was tiebreaking of how each action is preferred by each agent.
We proposed the hindrance metric and regret learning to easily improve the performance of PIBT with little additional computation.
Empirical results in both one-shot and lifelong MAPF reveal significant impacts of tiebreaking, proving that the proposed strategy is an attractive replacement for leading MAPF implementations.

The future direction includes adaptive construction of the PIBT preference during the planning.
This is motivated by our empirical results, which show that neither \hindrance nor \regret is always the best.
There are several realisations, such as selecting the best strategies through in-search learning or updating the weight parameters online, as have been studied for large neighbourhood search~\cite{phan2024adaptive}.
We also consider that directly optimising preferences with neural network policies is interesting~\cite{veerapaneni2024improving,jiang2024deploying}.
However, the inference with neural networks is notably slow compared to vanilla PIBT~\cite{skrynnik2024learn}, which impedes solving real-time and large-scale MAPF problems.
Therefore, we believe that further developments of heuristic-based preferences, as presented in this paper, are of practical value.

\section*{Acknowledgement}
This research was supported by a gift from Murata Machinery, Ltd.

\bibliography{sty/ref-macro,ref}
\end{document}